\documentclass[aps,prl,reprint,groupedaddress,superscriptaddress]{revtex4-2}

\usepackage[utf8]{inputenc}
\usepackage[T1]{fontenc}
\usepackage[german,english]{babel}

\usepackage{amsmath}
\usepackage{mathtools}
\usepackage{amssymb}
\usepackage{amsthm}
\usepackage{microtype} 
\usepackage{quoting} 
\quotingsetup{font=small}

\usepackage[pdftex]{graphicx}
\usepackage{fancyhdr}
\usepackage{bm}
\usepackage{comment}
\usepackage{booktabs}
\usepackage{eurosym} 
\usepackage{braket}
\usepackage{bbold} 
\usepackage{emptypage}
\usepackage{epstopdf}
\usepackage{verbatim}
\usepackage{enumitem}
\usepackage{float} 
\usepackage{chemformula}
\usepackage{isotope}

\usepackage{commath}

\usepackage{color}

\definecolor{Blue}{rgb}{0.00, 0.00, 1.00}
\definecolor{Red}{rgb}{1.00, 0.00, 0.00}

\begin{document}

\title{Emergent Bloch oscillations in a kinetically constrained Rydberg spin lattice}

\author{Matteo Magoni}
\affiliation{Institut für Theoretische Physik, Eberhard Karls Universität Tübingen, Auf der Morgenstelle 14, 72076 Tübingen, Germany}

\author{Paolo P. Mazza}
\affiliation{Institut für Theoretische Physik, Eberhard Karls Universität Tübingen, Auf der Morgenstelle 14, 72076 Tübingen, Germany}

\author{Igor Lesanovsky}
\affiliation{Institut für Theoretische Physik, Eberhard Karls Universität Tübingen, Auf der Morgenstelle 14, 72076 Tübingen, Germany}
\affiliation{School of Physics and Astronomy and Centre for the Mathematics and Theoretical Physics of Quantum Non-Equilibrium Systems, The University of Nottingham, Nottingham, NG7 2RD, United Kingdom}

\date{\today}

\begin{abstract}
We explore the relaxation dynamics of elementary spin clusters in a kinetically constrained spin system. Inspired by experiments with Rydberg lattice gases, we focus on the situation in which an excited spin leads to a ``facilitated'' excitation of a neighboring spin. We show that even weak interactions that extend beyond nearest neighbors can have a dramatic impact on the relaxation behavior: they generate a  linear potential, which under certain conditions leads to the onset of Bloch oscillations of spin clusters. These hinder the expansion of a cluster and more generally the relaxation of many-body states towards equilibrium. This shows that non-ergodic behavior in kinetically constrained systems may occur as a consequence of the interplay between reduced connectivity of many-body states and weak interparticle interactions. We furthermore show that the emergent Bloch oscillations identified here can be detected in experiment through measurements of the Rydberg atom density, and discuss how spin-orbit coupling between internal and external degrees of freedom of spin clusters can be used to control their relaxation behavior.
\end{abstract}






\maketitle

\textbf{Introduction.---} Kinetically constrained quantum systems have become an important setting for the investigation of complex dynamical many-body phenomena, both from the theoretical and the experimental point of view. In particular, constrained spin systems have turned out to constitute useful models for the study of slow relaxation, ergodicity breaking and the emergence of glassy physics \cite{LinMot,LinMot2,Turner2018,Scars2, Lesanovsky2012,Khemani2018,Shiraishi,SuraceMazza,Marcuzzi2016, Fac1, Fac2, LocFac, LocFac2, Fac3, PhysRevB.102.195150, Pancotti_2020}. In terms of experimental platforms a significant role is currently being played by Rydberg gases, in which atoms are excited to high-lying and strongly interacting states. This allows to implement effective quantum spin models with highly controllable state-dependent interactions that pave the way towards realizing a host of kinetic constraints \cite{Valado2016, 2020arXiv201007838L, Browaeys2020, Helmrich2020,Singer_2005, Pohl_2009, Aidelsburger:2015aa, Bloch_rev, Bloch2012, Jau2016, Bernien2017, Vibrational_ryd}.

Kinetic constraints impose restrictions on the connectivity between many-body states that break the Hilbert space into disconnected sectors~\cite{Ates2012,Lan2018,Ostmann2019a,PhysRevB.102.214437}. Ultimately, this may lead to the absence of thermalization and the emergence of non-ergodic behavior. This mechanism is different to ergodicity breaking stemming from disorder, occurring in many-body localized systems where it is caused by the emergence of local conservation laws \cite{MBL:Review}. Ergodicity breaking (in a disorder-free setting) may also occur when imposing externals fields: Refs. \cite{Rut1, Rutkevich_2018, Calabrese_nature, PPM, LeroseSurace, Konik1, konik2, Lagnese_2020, Pereira2020} show that for the case of a transverse field quantum Ising model, where an additionally applied longitudinal field leads to the confinement of excitations. This inhibits propagation of quasi-particles and thus prevents relaxation towards an ergodic steady state.

\begin{figure}[t!]
    \centering
    \includegraphics[width=0.8\linewidth]{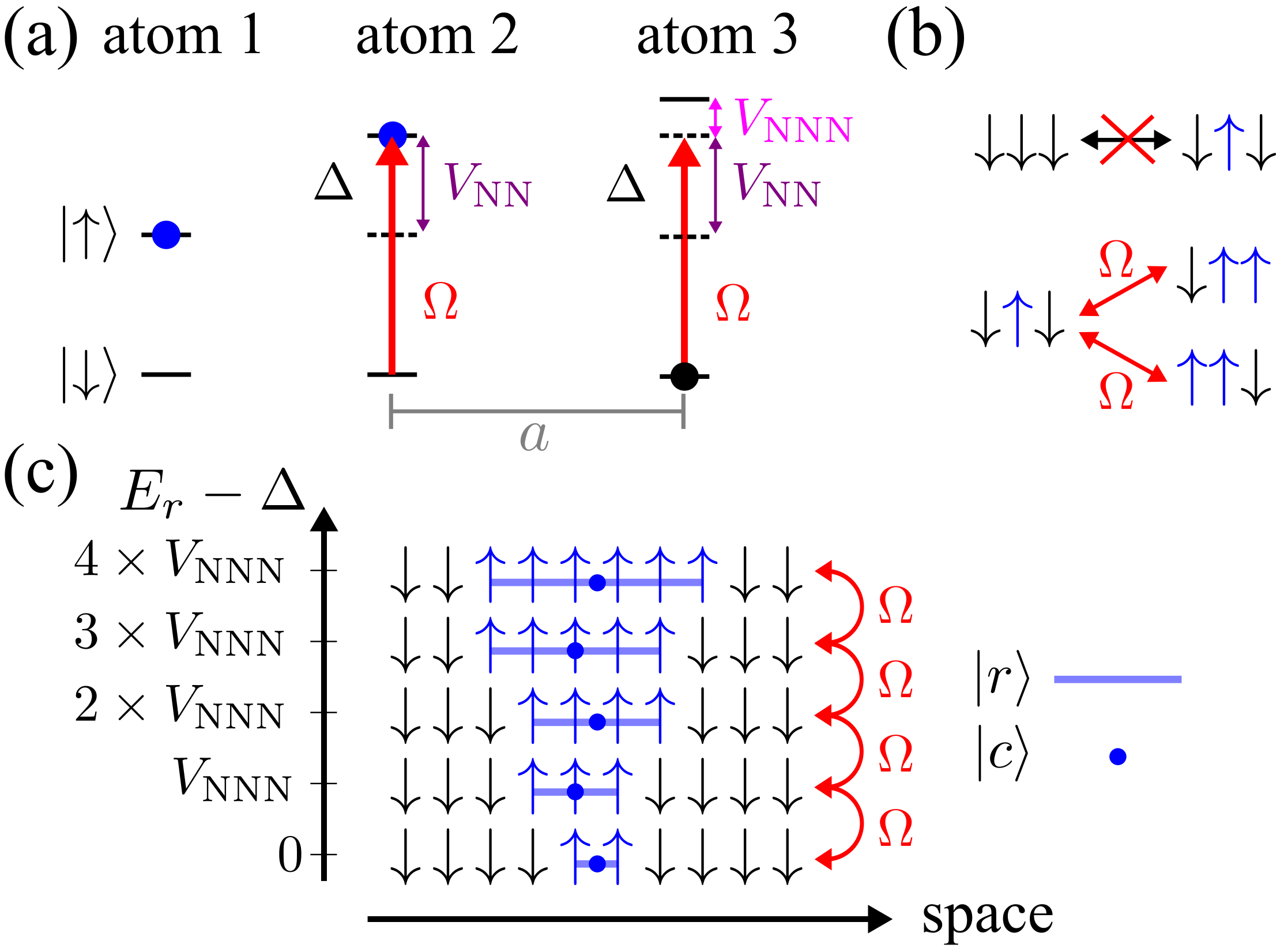}
    \caption{\textbf{Facilitated spin dynamics in a Rydberg lattice.} (a) Each atom is modeled as a two-level system, in which the states $\ket{\uparrow}$ and $\ket{\downarrow}$ represent the (excited) Rydberg state and the ground state, respectively. Atoms are separated by the distance $a$. $\Omega$ is the Rabi frequency of the excitation laser, which is detuned from the atomic transition by an energy $\Delta$. $V_{\text{NN}}$ and $V_{\text{NNN}}$ are the nearest neighbor and the next-nearest neighbour interaction strength between excited atoms. (b) In the facilitation regime a spin next to an excited spin is facilitated to (de)excite. This is achieved by setting $V_{\text{NN}} + \Delta = 0$. Each cluster can expand and shrink, but can neither merge with another cluster nor split. (c) Each cluster is parameterized by two coordinates: $\ket c$ (labeling the center of mass position) and $\ket r$ (labeling the number of excitations). The internal energy of a cluster of extension $r$ is given by $E_r=(r-2) V_{\text{NNN}}+\Delta$, generating a linear potential for spin clusters.}
    \label{fig:scheme}
\end{figure}
In this work we investigate the dynamics of a disorder-free, translationally invariant many-body quantum spin system under a so-called facilitation constraint. As shown in Fig. \ref{fig:scheme}, this can be realized with Rydberg atoms held in a lattice. We show that relaxation towards an ergodic stationary state is inhibited by the onset of Bloch oscillations of spin clusters --- composite states of domain wall quasi-particles --- which are caused by an emerging internal potential linear in the cluster size. These oscillations, which also have shown to herald non-ergodicity in the absence of disorder in different contexts, e.g. when external fields are imposed \cite{Calabrese_nature, LeroseSurace, van_Nieuwenburg_2019, Schulz_2019}, can be observed in the dynamics of the (Rydberg atom) density and thus are directly accessible in experiments. We argue that Bloch oscillations lead to an emergent ``Hilbert space fragmentation'' \cite{Sala_2020, Khemani_shattering, 2019arXiv191014048M}, reminiscent of fractonic systems \cite{Fractons, Pai_2020}. We furthermore show that there is a strong (spin-orbit) coupling between the internal dynamics of spin clusters and their external motion, which allows to construct either confined or propagating wave packets. Our work shows that constraints in conjunction with weak interactions offer new mechanisms for localization that go beyond merely breaking the Hilbert space connectivity of many-body states.

\textbf{Rydberg gas under facilitation conditions.---} We consider a one dimensional chain of $N$ atoms as depicted in Fig. \ref{fig:scheme}a. For each atom we employ a two-level description in terms of a fictitious spin-$\frac{1}{2}$ particle, which can be either in the up state $\ket{\uparrow}$ (excited Rydberg state) or in the down state $\ket{\downarrow}$ (ground state). Two atoms in the Rydberg state at neighboring sites have interaction energy $V_\mathrm{NN}>0$ (repulsive interactions). Including interactions up to next-nearest neighbors ($V_\mathrm{NNN}$), the Hamiltonian of the system is given by 
\begin{equation}
    H = \sum_{j=1}^N \left(\frac{\Omega}{2} \hat{\sigma}_j^x + \Delta \hat{n}_j + V_{\text{NN}} \, \hat{n}_j\hat{n}_{j+1} + V_{\text{NNN}} \, \hat{n}_j\hat{n}_{j+2} \right),
\label{eq:Hamilt}
\end{equation}
where $\Omega$ is the Rabi frequency and $\Delta$ is the laser detuning from the atomic transition energy. The operator $\hat{\sigma}^x=\left|\uparrow\right>\left<\downarrow\right|+\left|\downarrow\right>\left<\uparrow\right|$ effectuates transitions between the ground and Rydberg state, and $\hat{n}=\left|\uparrow\right>\left<\uparrow\right|$ projects on the Rydberg state. Periodic boundary conditions are also adopted.

We consider so-called facilitation conditions as depicted in Fig. \ref{fig:scheme}b. This means that the detuning cancels out the interaction between two adjacent atoms, i.e. $\Delta + V_{\text{NN}} = 0$. Moreover, we assume that the next-nearest neighbor interaction is small, i.e. $|V_{\text{NNN}}| \ll |\Delta|$. Under these conditions clusters of consecutive Rydberg excitations expand or shrink, but cannot (dis)appear, i.e. can neither split in two different clusters nor merge with another cluster. Hence the total number of clusters is a conserved quantity. Note, that this is rigorously true only when $\Delta \rightarrow \infty$ \cite{Abanin_DeRoeck, LeroseSurace}, and consequently we assume $\Delta$ to be the largest energy scale \footnote{See Supplemental Material for details on the experimental parameters needed to observe Bloch oscillations and the preparation of the initial states, which includes Refs. \cite{Trapping_frequency,Rydberg_lifetime,PairInteraction_software,Dressing_regime_gross,De_Leseleuc_2017}.}. For a single spin cluster a typical sequence of (near-)resonant transitions is e.g. given by $\ket{\downarrow \uparrow \downarrow \downarrow \downarrow \dots} \Leftrightarrow \ket{\downarrow \uparrow \uparrow \downarrow \downarrow \dots} \Leftrightarrow \ket{\downarrow \uparrow \uparrow \uparrow \downarrow \dots} \Leftrightarrow \ket{\downarrow \downarrow \uparrow \uparrow \downarrow \dots} \Leftrightarrow \ket{\downarrow \downarrow \uparrow \downarrow \downarrow \dots}$ \cite{Ostmann2019a}. It is thus convenient to describe the state of each cluster using two coordinates: the position of the center of mass (CM) $c$ and the number of excitations $r$ (see Fig.~\ref{fig:scheme}c). The internal energy of a cluster composed of $r$ excitations is then $E_r=r \Delta + (r-1) V_{\text{NN}} + (r-2) V_{\text{NNN}}=\Delta + (r-2) V_{\text{NNN}}$, where the facilitation condition was used in the last step. 

\textbf{Quasi-particle excitation spectrum.---} Let us begin from the situation in which a single cluster is present in a lattice of length $N$ (with the lattice spacing expressed in units of $a$) and assume periodic boundary conditions. It is then convenient to write the generic state of the cluster as a tensor product of the CM coordinate and the relative coordinate $\ket{\psi} = \ket c \otimes \ket r$, where $c$ is an index that labels the position of the CM of the cluster and $r$ denotes the number of excitations. The total number of possible values of the CM coordinate is thus $2N$: $c \in \{\frac{1}{2}, 1, \frac{3}{2}, ..., N\}$. Integer values of $c$ refer to on-site CMs, while the half-integer values correspond to CMs located at the midpoints between two lattice sites. The coordinate $r$ is instead an integer number between $1$ and $N-1$, since a cluster with $N$ excitations is not allowed. Thus, for instance, $\ket{2}\ket{3} = \ket{\uparrow \uparrow \uparrow \downarrow \downarrow \dots}$ and $\Ket{\frac{5}{2}}\ket{2} = \ket{\downarrow \uparrow \uparrow \downarrow \downarrow \dots}$. 

Given this representation, there are four possible transitions that a state $\ket c \ket r$ can undertake (at rate $\Omega$), provided that $1 < r < N-1$ (when $r=1$ the cluster can only increase, when $r=N-1$ the cluster can only decrease). Possible target states are (see Fig. \ref{fig:scheme}): $\Ket{c + \frac{1}{2}} \ket{r+1}$ (the spin to the right of the rightmost excitation flips up), $\Ket{c - \frac{1}{2}} \ket{r+1}$ (the spin to the left of the leftmost excitation flips up), $\Ket {c - \frac{1}{2}} \ket{r-1}$ (the rightmost excitation flips down), $\Ket{c + \frac{1}{2}} \ket{r-1}$ (the leftmost excitation flips down). Note, that these transitions rules, which determine the kinetic energy of a spin cluster, imply a spin-orbit coupling, i.e. a coupling between the (internal) relative coordinate of the cluster and the (external) CM dynamics. This is because a cluster cannot shrink/expand without changing its CM position. Taking furthermore into account the internal energy, $E_r$, which only depends on the cluster length $r$, the effective Hamiltonian describing a single spin cluster reads
\begin{eqnarray}
H &= & \Omega \sum_{c=\frac{1}{2}}^{N} \sum_{r=1}^{N-2} \left[\Ket{c+\tfrac{1}{2}} \bra c \otimes \left(\ket{r+1}\bra{r} + \text{h.c.} \right) + \text{h.c.} \right]  \nonumber \\
&&+ V_{\text{NNN}} \sum_{c=\frac{1}{2}}^{N} \sum_{r=2}^{N-1} (r-2) \ket c \bra c \otimes \ket r \bra r.
\label{eq:full_hamiltonian}
\end{eqnarray}
Taking the Fourier transform of the CM coordinate $\ket c = \frac{1}{\sqrt{2N}} \sum_q e^{iqc} \ket q$, where $q=-2\pi + \frac{2\pi}{N}k$ with $k=0,1,\dots,2N-1$, simplifies the Hamiltonian to $H = \sum_q H_q \ket q \bra q$, with
\begin{eqnarray}
H_q &=& 2 \Omega \cos\left(\frac{q}{2}\right) \sum_{r=1}^{N-2} \left(\ket{r+1} \bra r + \text{h.c.} \right) \nonumber\\
&&+ V_{\text{NNN}}\sum_{r=2}^{N-1} (r-2) \ket r \bra r.
\label{eq:H_q}
\end{eqnarray}
Note, that the CM momentum quantum number, $q$, takes on values between $-2\pi$ and $2 \pi$ because of the $2N$ lattice sites that make up the CM lattice. The Hamiltonian $H_q$ of each CM momentum sector can be interpreted as that of a particle hopping with a rate $J_q=2 \Omega \cos\left(\frac{q}{2}\right)$ through a semi-infinite lattice, subject to a linear potential of slope $V_\mathrm{NNN}$. This potential, however, affects only "sites" with coordinate $r>2$. 

\begin{figure}[t]
    \centering
    \includegraphics[width=\linewidth]{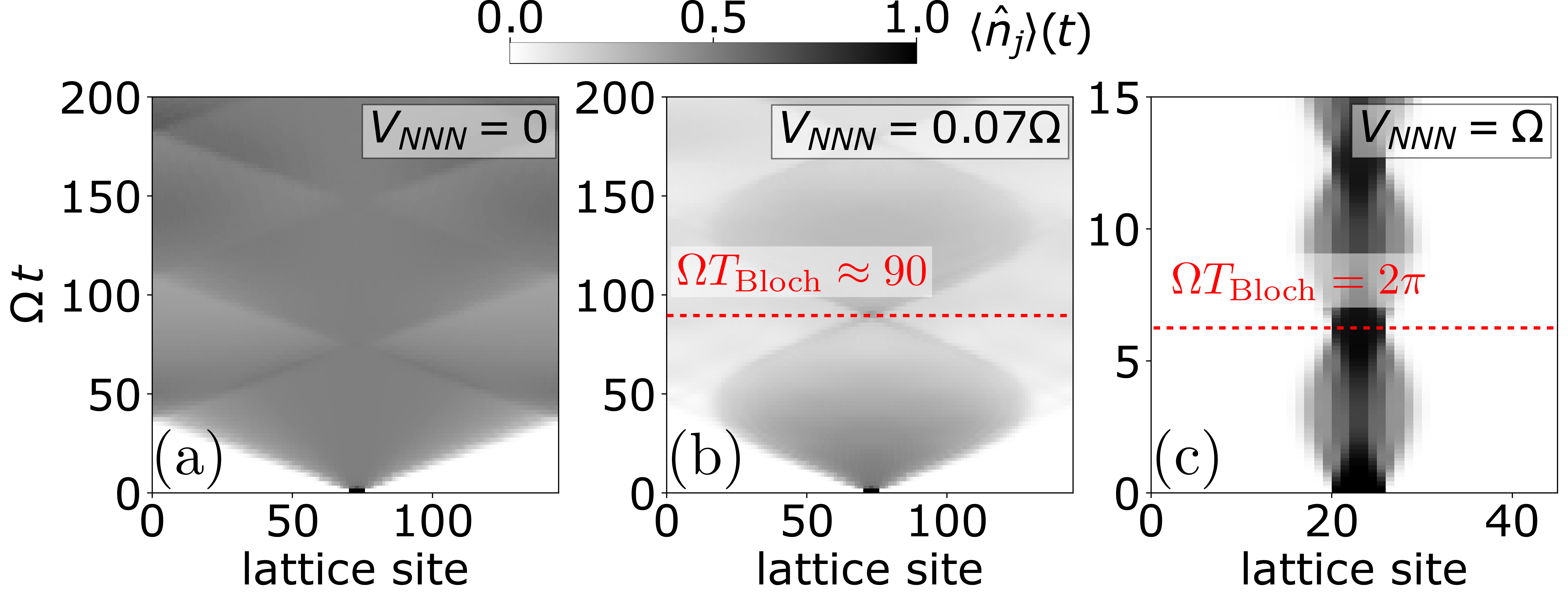}
    \caption{\textbf{Time evolution of a spin cluster.} Shown is the local Rydberg excitation density $\left\langle \hat{n}_j\right\rangle (t)$. (a) Ballistic expansion of spin cluster. (b) Coexistence of Bloch oscillations and ballistic expansion. (c) Bloch oscillations with period $T_\text{Bloch} = 2 \pi/ V_{\text{NNN}}$. The initial state is $\ket{\psi(0)} = \ket{c_0} \otimes \ket{r_0}$ with $r_0 = 6$ and $c_0$ chosen appropriately for each panel.}
    \label{fig:bloch_examples}
\end{figure}
\textbf{Spin cluster Bloch oscillations.---} Evidently, there is a close connection between the spin cluster dynamics and that of a hopping particle in a tilted lattice. The latter is a classic problem in condensed matter physics \cite{Bloch1929, Zener, Wannier1960, Wannier1962, Fogedby1973, Hartmann_2004}, and one of its most striking dynamical features is the emergence of so-called Bloch oscillations. Our situation bears some differences, though, such as the semi-infinite lattice and the fact that spin clusters are composite objects that have an internal and an external structure. In the following we consider the dynamics of a spin cluster that is initially prepared (time $t=0$) in a state with CM position $c_0$ and contains $r_0$ consecutive excitations: $\ket{\psi(0)} = \ket{c_0} \otimes \ket{r_0}$ (note, that $c_0$ and $r_0$ have to be compatible, e.g. when $r_0$ is odd, $c_0$ has to be integer). This state evolves according to 
\begin{equation}
\ket{\psi(t)} = e^{-i Ht}\ket{\psi(0)} = \frac{1}{\sqrt{2N}} \sum_q e^{iqc_0} \ket{q} \otimes e^{-i H_q t} \ket{r_0} \label{eq:initial_flat_state}.
\end{equation}
In Fig. \ref{fig:bloch_examples} we show the site-resolved Rydberg excitation density --- a quantity which can be experimentally measured \cite{Browaeys2020} --- for a spin cluster of initial size $r_0=6$. In the absence of next-nearest neighbor interactions this cluster expands linearly in time and the density shows the expected light cone structure. For large $V_\mathrm{NNN}$, however, we see unambiguously Bloch oscillations of the spin cluster size, whose period is given by $T_{\text{Bloch}} = 2 \pi/V_{\text{NNN}}$. At intermediate values of $V_\mathrm{NNN}$ we observe that Bloch oscillations and ballistic expansion coexist. The reason for this behavior is the composite nature of the spin domain excitation together with the fact that the tilted lattice is actually semi-infinite. As can be seen from Hamiltonian (\ref{eq:H_q}) each $q$-component is governed by a different "hopping rate" $J_q$. For a given hopping rate the amplitude of the Bloch oscillations is then $l_{\text{Bloch}} \simeq J_q/V_\mathrm{NNN}$. This relation, however, holds only in case of an infinite lattice. For a spin cluster to effectively experience such infinite lattice its initial size $r_0$ must be larger than $l_{\text{Bloch}}$, so that Bloch oscillations never reach its boundary. For the parameters chosen for Fig. \ref{fig:bloch_examples}b this is, however, true only for certain values of the CM momentum $q$ on which the initial state has support. Therefore, these components perform Bloch oscillations while the others expand ballistically. This results in the coexistence behavior displayed in the panel. 

Requiring that the initial cluster size $r_0$ is large enough, such that no $q$-component of the state experiences the edge of the lattice, defines a lower threshold for the next-nearest neighbor interaction:
\begin{equation}
V_{\text{NNN}} \gtrsim \frac{2 \Omega}{r_0}.
\label{eq:condition}
\end{equation}
If this condition is satisfied, then perfect Bloch oscillations as shown in Fig. \ref{fig:bloch_examples}c are observed.
\begin{figure}[t]
    \centering
    \includegraphics[width=\linewidth]{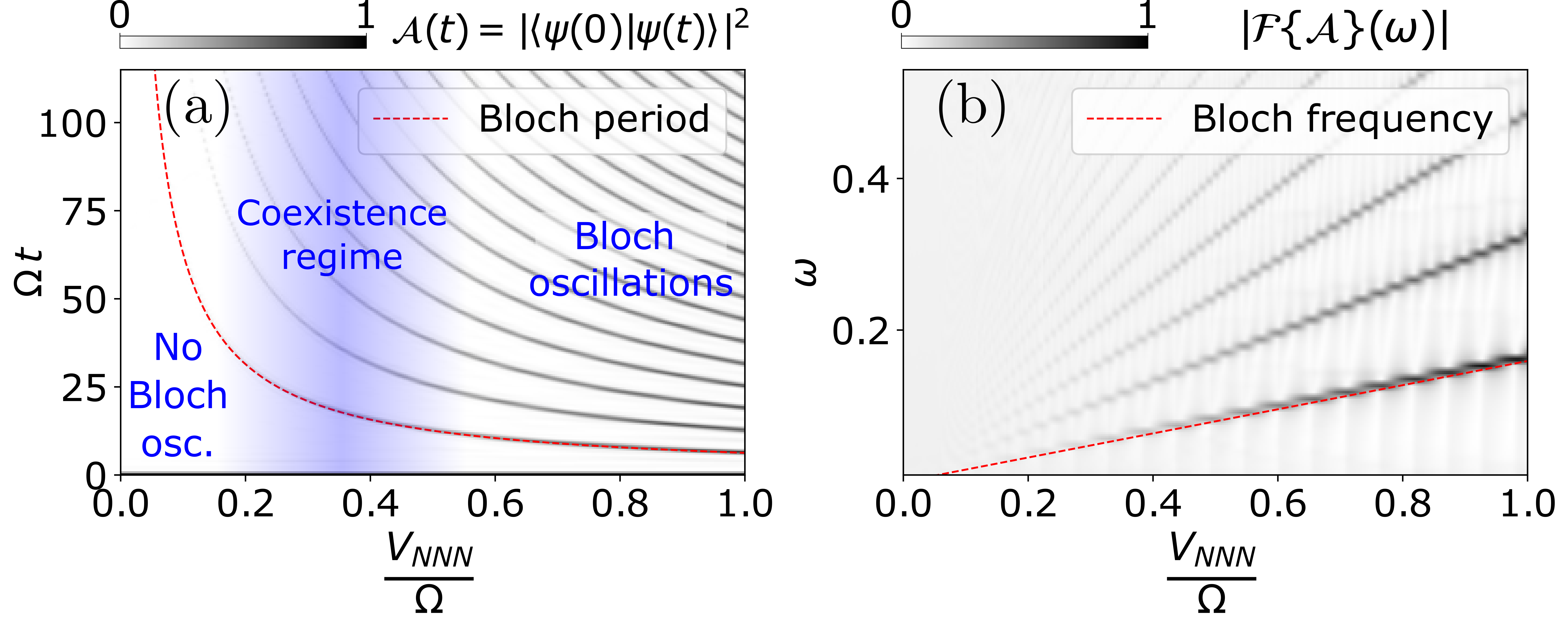}
    \caption{\textbf{Auto-correlation function and its Fourier transform.} (a) Auto-correlation function, Eq. (\ref{eq:auto_c_function}), as a function of the next-nearest neighbor interaction strength $V_{\text{NNN}}$. The red dashed line shows the expected Bloch oscillation period, $T_{\text{Bloch}} = 2 \pi/V_{\text{NNN}}$. Bloch oscillations become visible through marked peaks only for sufficiently large $V_{\text{NNN}}$. In the coexistence regime some wave packet components undergo ballistic expansion. (b) Fourier transform of the auto-correlation function. The red dashed line shows the expected Bloch oscillation frequency.}
    \label{fig:autocorrelation}
\end{figure}
To quantify the onset of Bloch oscillations and the concomitant periodic behaviour, we define the auto-correlation function 
\begin{equation}
\mathcal{A}(t) = |\braket{\psi(0)|\psi(t)}|^2, \label{eq:auto_c_function}
\end{equation}
which measures the overlap between the initial state and the state at time $t$. Fig.~\ref{fig:autocorrelation}a shows that there is a gradual passage from a regime of ballistic expansion to one with Bloch oscillations with an intermediate coexistence regime, indicated with the blue shaded area. As expected, one observes pronounced time-periodic behavior for sufficiently large values of $V_\mathrm{NNN}$, with high-amplitude revivals at the Bloch period $T_{\text{Bloch}} = 2 \pi/V_{\text{NNN}}$. Decreasing the next-nearest neighbor interaction strength reduces the amplitude of these revivals, which is due to certain CM momentum components evolving ballistically. This behavior is also reflected in the Fourier transform of the auto-correlation function shown in Fig.~\ref{fig:autocorrelation}b, which displays a clear peak at the Bloch frequency (and higher harmonics) up to a threshold value of $V_\mathrm{NNN}$.

\begin{figure}[t]
    \centering
    \includegraphics[width=\linewidth]{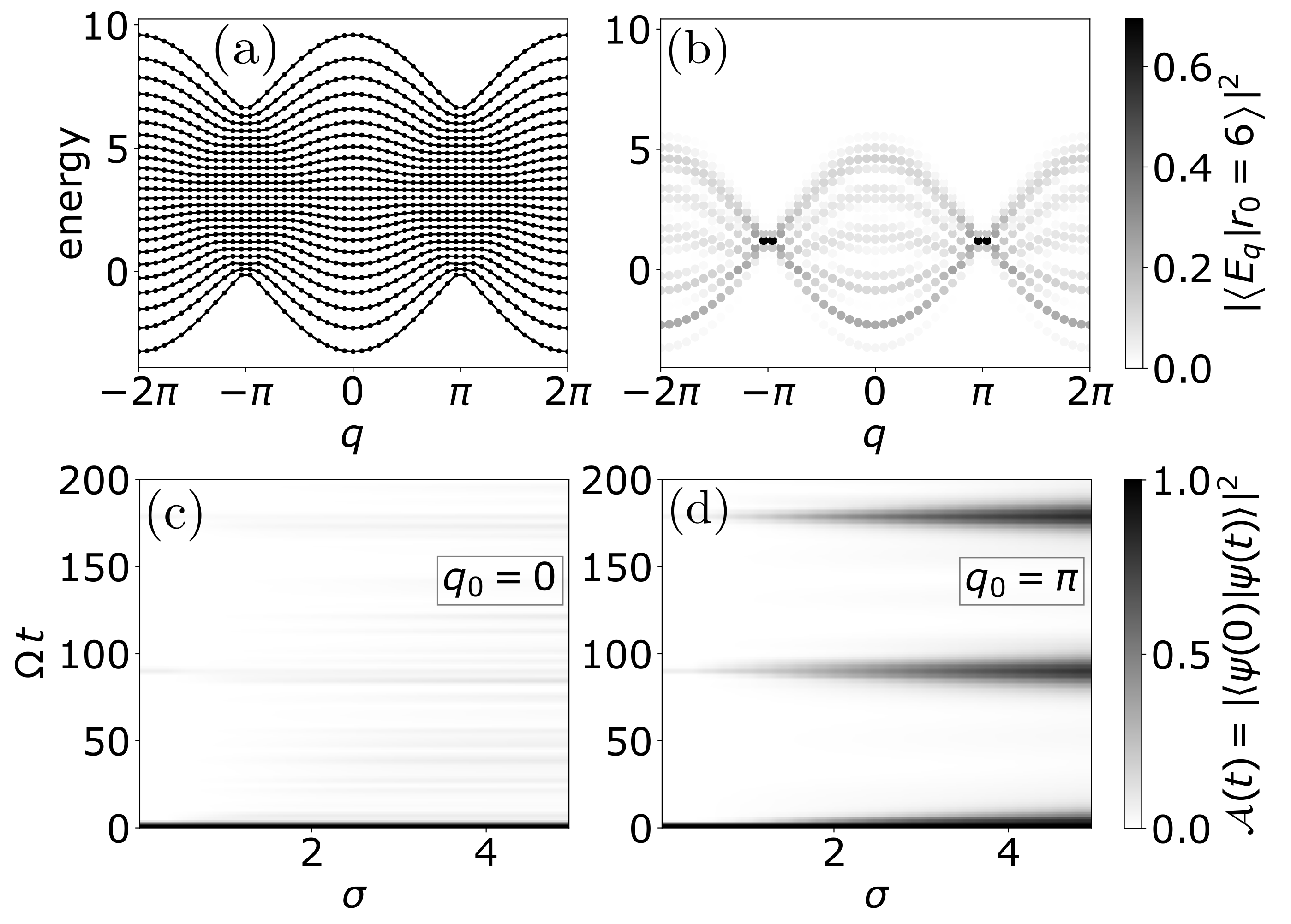}
    \caption{\textbf{Band structure and auto-correlation function}. (a) Band structure given by the Hamiltonian $H_q$. Due to the presence of the linear potential introduced by the next-nearest neighbor interaction, the bands are locally flat and non-degenerate at $q = \pm \pi$. (b) Projection of the state $\ket{r_0 = 6}$ on the eigenstates of $H_q$, for different $q$. The overlap is larger for the low-energy eigenstates, and at $q = \pm \pi$ the state $\ket{r_0 = 6}$ is an eigenstate. (c) Auto-correlation function for an initial state centered on $q_0 = 0$. At $t=0$ it has the value $1$ and then decays rapidly. No periodic behavior is visible as $\sigma$ increases. The faint features are due to finite size effects. (d) Auto-correlation function for an initial state centered on $q_0 = \pi$. Bloch oscillations appear as $\sigma$ increases. In (c) and (d) we choose $V_\mathrm{NNN} = 0.07 \, \Omega$ and $r_0 = 6$, which corresponds to the situation shown in Fig.~\ref{fig:bloch_examples}b.}
    \label{fig:fig4}
\end{figure}

\textbf{Spin-orbit coupling of spin clusters.---} So far we have analyzed the evolution of an initial state that is composed of a single spin cluster with fixed size $r_0$ and CM position $c_0$. As can be seen from Eq. (\ref{eq:initial_flat_state}) such state has an equal weight (up to a phase factor) on all $q$-components. This state is certainly the one that is most naturally prepared in experiment \cite{Note1}. However, to fully investigate how the (spin-orbit) coupling between the relative and CM motion affects the dynamics of spin clusters, it is instructive to study initial states in which the CM position is not fixed, but has the form of a Gaussian wave packet of finite width $\sigma$:
\begin{equation}
\ket{\psi(0)} = \left(\frac{1}{\mathcal{N}} \sum_{c} e^{-\frac{(c-c_0)^2}{4 \sigma^2}} e^{-i q_0 c} \ket c \right) \otimes \ket{r_0}.
\label{eq:superposition}
\end{equation}
Here $\mathcal{N}$ is a normalization constant and $c_0$ and $q_0$ are the average CM position and momentum. A possible protocol for creating such state is discussed in \cite{surace2021scattering}. In order to get a first idea of what dynamical behavior to expect, we consider the band structure of the Hamiltonian $H_q$ [Eq. (\ref{eq:H_q})], which is depicted in Fig.~\ref{fig:fig4}a. One observes a series of $N-1$ bands that govern the dynamics of the relative motion (spin cluster expansion and reduction) as a function of the CM momentum $q$, which is a direct manifestation of the coupling between those two degrees of freedom. Note, that in case of an infinite lattice the bands would be flat and equally spaced forming the well known Wannier-Stark ladder \cite{Wannier1960}. In a finite lattice, the eigenvalues of each Hamiltonian $H_q$ are instead given by the zeros of the $q$-dependent Lommel polynomial of degree $N-1$ \cite{Stey_1973}. At $q= 0, \pm \pi, \pm 2\pi $ the gradient of the bands is zero and the group velocity of a wave packet centred here vanishes. These points are thus well-suited for an analysis in terms of the auto-correlation function which otherwise would decay in time simply due to the linear motion of the wave packet. At $q=\pm\pi$ the hopping term in Eq. (\ref{eq:H_q}) vanishes and the Hamiltonian becomes diagonal in the $\ket r$ basis. This is seen in Fig. \ref{fig:fig4}b which shows the projection of the initial spin domain state $\ket{r_0}$ (here with $r_0=6$) on the eigenstates of $H_q$. The plot also shows that for all other values of $q$ the state $\ket{r_0}$ is not among the eigenstates but is formed by a superposition of them.

In Figs.~\ref{fig:fig4}c,d we show the auto-correlation function using the initial state (\ref{eq:superposition}) with $q_0 = 0$ and $q_0 = \pi$, respectively, as a function of $\sigma$. Both plots are obtained with the parameters of Fig.~\ref{fig:bloch_examples}b. For $\sigma=0$ both situations correspond (up to a phase factor) to the one described by the state (\ref{eq:initial_flat_state}), i.e. we expect to observe a coexistence between Bloch oscillations and ballistic expansion. Indeed, here both panels show the same faint signatures of Bloch oscillations. Upon an increase of $\sigma$ these are amplified for $q_0 = \pi$ and vanish for $q_0 = 0$. When $q_0 = \pi$, the initial state has support on the values of $q$ that minimize the width of Bloch oscillations. Therefore the initial spin cluster size $r_0$ is large enough not to see the boundary of the semi-infinite lattice. Conversely, when $q_0 = 0$, the width of the oscillations exceeds the initial cluster width $r_0$ and no Bloch oscillations appear. This result suggests that, for any given values of $V_\mathrm{NNN}$ and $r_0$, it is always possible to engineer an initial wave function sufficiently peaked around $q_0 = \pi$ that undergoes a periodic dynamics. This is a direct consequence of the spin-orbit coupling: Bloch oscillations are dynamical features of the relative spin cluster dynamics, but they can be controlled by the selection of specific CM momenta $q$.

\textbf{Absence of relaxation in the many-body system.---} Having understood the dynamics of a single spin cluster allows to make statements on the dynamics in the many-body case: a general initial state can be decomposed into basis states containing $m$ clusters. These shall be labeled with the coordinates $\ket{c_i}$ and $\ket{r_i}$, for $i=1,2,\dots,m$. Such a state does not relax when each of the individual clusters performs Bloch oscillations without "touching" neighboring clusters. Such situation occurs when each of the $m$ clusters has a length of at least $r_0 \simeq 2 \Omega \big/ V_{\text{NNN}}$, and when the distance between any two neighboring clusters is also at least $r_0$. The first requirement derives from Eq.~(\ref{eq:condition}) and ensures the emergence of Bloch oscillations within all $m$ clusters. The second requirement comes from the fact that two neighboring clusters must not meet when oscillating with amplitude $l_\text{Bloch}$. A lower bound, $\Gamma(N)$, for the number of many-body basis states that satisfy both conditions can be derived from the number $W(m)$ of ways $m$ hard rods of length $2 r_0$ can be arranged in a lattice of length $N$. With $W(m) = [(N-2 r_0 m+m)!]/[(N-2r_0 m)! \, m!]$ and using that at most $N/(2r_0)$ rods can be inserted in a system of length $N$, we find that $\Gamma(N) = \sum_{m=1}^{\frac{N}{2r_0}} W(m)\sim C^N$, with $1<C<2$. Thus, the number of many-body states, whose dynamics is frozen due to spin cluster Bloch oscillations, scales exponentially with the lattice size. Therefore, non-relaxing states are not rare in Hilbert space. This result is closely related to the fragmentation of Hilbert space observed in fractonic systems \cite{Fractons}, characterized by a restricted mobility of elementary excitations. However, differently from fractonic models, in our system there are no notions of charge and dipole moment conservation, and the emergence of disconnected Hilbert subspaces results from the restricted mobility of composite spin clusters. 

\textbf{Conclusions and outlook.---} Observations related to ours have been made in quantum Ising chain systems~\cite{Calabrese_nature, van_Nieuwenburg_2019, Schulz_2019}, when an applied external longitudinal field penalises the creation of extended spin clusters. In our case, however, confinement is created by interactions within the spin system itself, leading to emergent Bloch oscillations, whose dynamics is strongly dependent by the coupling between the internal and external dynamics of spin clusters. This shows that even weak interactions within constrained systems can have a dramatic impact on the ability to relax. In the future, it would be interesting to develop a scattering theory that describes collisions between two spin clusters, and to generalize the study to kinetically constrained spin systems in two dimensions.

\acknowledgements We acknowledge support from EPSRC (Grant No. EP/R04421X/1), the  “Wissenschaftler-Rückkehrprogramm GSO/CZS” of the Carl-Zeiss-Stiftung and the German Scholars Organization e.V., as well  as  through  the  Deutsche  Forschungsgemeinsschaft (DFG, German Research Foundation) under Project No. 428276754, through SPP 1929 (GiRyd), and under Germany’s Excellence Strategy - EXC No. 2064/1 - Project No. 390727645.

\bibliography{bib}

\clearpage

\onecolumngrid
\begin{center}
{\large{\textbf{SUPPLEMENTAL MATERIAL}}}
\end{center}

\vspace{11pt}

\setcounter{equation}{0}
\setcounter{figure}{0}
\setcounter{table}{0}
\renewcommand{\theequation}{S\arabic{equation}}
\renewcommand{\thefigure}{S\arabic{figure}}
\renewcommand{\bibnumfmt}[1]{[#1]}
\renewcommand{\citenumfont}[1]{#1}
\renewcommand{\theequation}{S\arabic{equation}}
\renewcommand{\thefigure}{S\arabic{figure}}
\renewcommand{\bibnumfmt}[1]{[#1]}
\renewcommand{\citenumfont}[1]{#1}

We present a discussion on a possible experimental procedure to detect Bloch oscillations in Rydberg quantum simulators. In the first section we provide values for the experimental parameters and energy scales needed to observe Bloch oscillations, together with an estimate of their period and the comparison with the lifetime of the considered Rydberg state. In the second section, we elaborate on the preparation of the initial state presented in the main text.

\section{Experimental considerations}
We focus here on \isotope[87]{Rb} atoms. However, experiments conducted with other atoms like \isotope[39]{K} or \isotope[133]{Cs} feature parameters of similar order of magnitude. In current experimental platforms, atoms are typically held at a distance $a \simeq 5 \mu m$ in optical traps with trapping frequency $\omega \simeq 100$ kHz \cite{Trapping_frequency}. The lifetime of \isotope[87]{Rb} Rydberg excitations with principal quantum number $n = 70$ in the $s$ orbital and at temperature $T = 300 \, \mathrm{K}$ is $\tau_\mathrm{life} \simeq 1.5 \cdot 10^{-4} s$ \cite{Rydberg_lifetime}. The anti-blockade constraint prescribes that the laser detuning cancels out the nearest-neighbor interaction, i.e. $\Delta + V_\mathrm{NN} = 0$. In presence of van der Waals interaction, the interaction between nearest-neighbor Rydberg excitations is $V_\mathrm{NN} = \frac{C_6[ns]}{a^6}$. The $C_6$ coefficient is approximately proportional to $n^{11}$, where $n$ is the principal quantum number of the Rydberg state. For $n = 70$, $C_6[ns] = 862.8$ GHz $\mu m^6$ \cite{PairInteraction_software} and therefore $V_\mathrm{NN} = 55.2$ MHz. In order to satisfy the anti-blockade constraint, this must be also the value of the detuning, i.e. $\Delta = 2\pi \cdot 8.8$ MHz.

The next-nearest-neighbor interaction is given by $V_\mathrm{NNN} = \frac{C_6[70s]}{(2a)^6} = 862.8$ kHz. In order to observe Bloch oscillations we have to satisfy Eq. (5) of the main text, that constrains the possible values for the Rabi frequency. For an initial cluster of, say, $r_0 = 6$ excitations, the Rabi frequency must then be $\Omega \lesssim 2\pi \cdot 412$ kHz. This shows that both the facilitation constraint and Eq. (5) are simultaneously satisfied in the so called dressing regime \cite{Dressing_regime_gross}, when the laser detuning is much larger than the Rabi frequency.

The period of Bloch oscillation is given by $T_\mathrm{Bloch} = \frac{2 \pi}{V_\mathrm{NNN}} = 7.18 \cdot 10^{-6} s$. This means that within the lifetime $\tau_\mathrm{life}=1.5 \cdot 10^{-4} s $ of the Rydberg excitations it is possible to observe 20 Bloch oscillations. This shows that Rydberg atom platforms currently implemented in laboratory can access the range of parameters necessary to observe real-space Bloch oscillations.

\section{Preparation of the initial state}

As highlighted in the main text, the structure of the initial state strongly affects the subsequent dynamics. Here we elaborate on the possible procedure to prepare the two initial states presented in the main text. 
The preparation of the initial state
\begin{equation}
    \ket{\psi(0)} = \ket{c_0} \otimes \ket{r_0}
    \label{eq:initial state}
\end{equation}
with a fixed CM position and defined number of Rydberg excitations is feasible in experiments that permit single site addressability, such as \cite{Bernien2017} and \cite{De_Leseleuc_2017}. This allows to change the state of individual atoms to the Rydberg state and thus may be utilized to prepare clusters of consecutive Rydberg excitations.

The preparation of the initial state Eq. (7), being a Gaussian superposition of clusters with fixed number of excitations but different CM positions, is certainly more challenging. However, a similar problem has been discussed in \cite{surace2021scattering}. In this work - which considers the dynamics of a quantum Ising model - spatial inhomogeneities are exploited to create a spatially locally modified band structure for elementary excitations. This procedure relies on the fact that when an excitation, while spreading ballistically, enters this spatial region, only certain momenta components are transmitted forward. This mechanism narrows the transmitted wave packet in momentum space. Thus, this protocol realizes a filter that can select momentum components from an initial state that is entirely delocalized in momentum space, such as Eq. (\ref{eq:initial state}). 

It should be possible to implement a similar protocol in the scenario considered in our manuscript: an initial state with fixed CM position like Eq.~(\ref{eq:initial state}) is localized in real space and thus entirely delocalized in momentum space. This means that all the momenta are equally populated. To create an initial state with better defined central momentum one can introduce an inhomogeneity in the lattice for example by varying locally the lattice spacing. This leads to a locally changing band structure that acts as a filter for momentum components. During the passage through this inhomogeneity only some momenta are transmitted, leading to a scattered/transmitted wave packet with a reduced support on the CM momenta.

\end{document}